\documentclass[onecolumn,prd,superscriptaddress,nofootinbib,notitlepage]{revtex4-1}
\pdfoutput=1
\usepackage[T1]{fontenc}
\usepackage{bm}
\usepackage{subfigure}
\usepackage{amssymb}
\usepackage{graphicx}
\usepackage{amsmath}
\usepackage{tensor}
\usepackage{epstopdf}
\usepackage{extarrows}
\usepackage{xcolor}
\usepackage[colorlinks=true,
              linkcolor=blue,
              urlcolor=blue,
              citecolor=blue
              ]{hyperref}

\newcommand{\be}{\begin{equation}}
\newcommand{\ee}{\end{equation}}
\newcommand{\bea}{\setlength\arraycolsep{2pt} \begin{eqnarray}}
\newcommand{\eea}{\end{eqnarray}}

\def\0{{\sst{(0)}}}
\def\1{{\sst{(1)}}}
\def\2{{\sst{(2)}}}
\def\3{{\sst{(3)}}}
\def\4{{\sst{(4)}}}
\def\5{{\sst{(5)}}}
\def\6{{\sst{(6)}}}
\def\7{{\sst{(7)}}}
\def\8{{\sst{(8)}}}
\def\sst#1{{\scriptscriptstyle #1}}

\begin{document}

\title{Topological Interpretation of Black Hole Phase Transition in Gauss-Bonnet Gravity}

\author{C. Fairoos}
\email{fairoos.phy@gmail.com}
\affiliation{Department of Physics, T. K. M. College of Arts and Science Kollam, India- 691005 }

\begin{abstract}

Phase transitions of Einstein-Gauss-Bonnet black holes are studied using Duan's $\phi-$ field topological current theory, where black holes are treated as topological defects in the thermodynamic parameter space. The kinetics of thermodynamic defects are studied using Duan's bifurcation theory. In this picture, a first-order phase transition between small/large black hole phases is interpreted as the interchange of winding numbers between the defects as a result of some action at a distance. We observe a first-order phase transition between small/large black holes for  $D=5$  {Einstein-}Gauss-Bonnet theory similar to Reissner-Nordstr\"{o}m black holes in AdS space. This implies that these black hole solutions share the same topology and phase structure. We have also studied the phase transition of neutral black holes in $D\geq 6$ and found a transition between unstable small and large stable black hole phases similar to the case of neutral black holes in AdS space. Recently, it has been conjectured that black holes with similar topological structure exhibit the same thermodynamic properties. Our results strengthen the conjecture by connecting the topological nature of black holes to phase transitions.

\end{abstract}
\maketitle

\section {Introduction}
Statistical and thermodynamic studies of black holes are helpful for a better understanding of gravity. A major goal in this aspect is to develop a theory that can explain both the large-scale behaviour of the universe and the small-scale behaviour of the matter. Black holes in asymptotic anti-de Sitter (AdS) space is a promising candidate to explore thermodynamic properties due to many reasons. These black hole spacetimes can be described by a dual thermal field theory and accordingly, a microscopic description of underlying degrees of freedom can be obtained \cite{Maldacena:1997re,Gubser:1998bc, Witten:1998qj}. It turns out that this gauge/gravity duality is a strong tool in many systems such as quark-gluon plasma, high temperature superconductors and condensed matter systems. Also, black holes in AdS space were shown to possess rich phase structures. \\

It was Hawking and Page who demonstrated the existence of a certain phase transition between thermal radiation and black hole in the phase space of Schwarzschild-AdS black hole for the first time\cite{Hawking:1982dh}. Later, a charged black hole in AdS spacetime was shown to exhibit first order phase transition between small/large black hole phases similar to van der Waals-Maxwell liquid-gas system\cite{Chamblin:1999tk,Chamblin:1999hg}. Consequently, the critical behaviour and phase transition of various black hole solutions were studied in the extended phase space where the cosmological constant ($\Lambda$) serves the role of thermodynamic pressure and its conjugate parameter as thermodynamic volume \cite{Kubiznak:2012wp, Cai:2013qga, Wei:2012ui, Xu:2015rfa, Dayyani:2017fuz,Wei:2015iwa}. \\

Thermodynamic properties of black holes can be deduced using a phenomenological theory of topology. According to this, black holes are identified as topological defects of thermodynamic parameter space. Moreover, it has been conjectured in \cite{Wei:2022dzw} that all black hole solutions can be classified into three groups based on their topological charge or number ($W$), i.e., $W=-1, 0, 1$, where the topological charge associated with the parametric space of black hole is obtained using Duan's $\phi$- field topological current theory \cite{Duan:1984ws}. This conjecture is verified for black holes in various spacetimes including modified gravity theories \cite{Yerra:2022alz,Yerra:2022coh, Bai:2022klw, Wu:2022whe, Fang:2022rsb, Liu:2022aqt, Du:2023wwg, Fairoos:2023jvw,Du:2023nkr,Wu:2023sue,Wu:2023xpq,Wu:2023meo}. The black hole solutions with similar topological nature are expected to show similar thermodynamic properties and are classified accordingly. Besides black hole systems, the topological current theory has been used in diverse areas in theoretical physics such as cosmology and condensed matter physics\cite{Du:1997hel,Du:1998ygd}.\\

The topology of black hole {thermodynamics} can be further explored to study phase transitions using Duan's bifurcation theory of topological current\cite{Duan:1998kw}. Accordingly, the black hole phase transition can be viewed as a once or twice interchange of winding numbers between topological defects \cite{Fan:2022bsq}. In this paper we try to interpret black hole phase transitions in Einstein-Gauss-Bonnet {(EGB)} gravity theory in terms of topology.  Earlier studies show that a charged {EGBAdS} black hole, Reissner–Nordström (RN)-AdS black holes in arbitrary dimensions, and neutral {EGB}AdS black holes in 5D belong to the same topological class with topological number $W=+1$. Similarly, neutral {EGB}AdS black holes in $D\geq 6$ and 4D neutral black holes in AdS space share the same topology class with $W=0$ \cite{Wei:2022dzw,Liu:2022aqt}. We explicitly verify if the topological classification of black hole solutions is valid for phase transition properties as well.\\

The structure of the paper is as follows: in \ref{TD}, we briefly summarize the thermodynamics of black holes in {EGB}AdS spacetime. In \ref{TP}, we discuss Duan's $\phi$- field topological current theory and bifurcation theory. The interpretation of phase transition in terms of topology is presented for 5D {EGB} black holes in \ref{5D} and higher dimensional holes in \ref{HD}. Our results are concluded in \ref{FL}.\\

\section{Thermodynamics of Gauss-Bonnet Black Holes in AdS Space}\label{TD}
We briefly review the  black hole thermodynamics in $D$ dimensional {EGB} gravity theory governed by the action,
\bea \nonumber
S=\frac{1}{16 \pi} \int d^Dx \sqrt{-g}\Bigg[ R - 2 \Lambda + \alpha_{GB}\Big[R_{\mu \nu \rho \sigma} R^{\mu \nu \rho \sigma} \\ 
- 4 R_{\mu \nu} R^{\mu \nu} + R^2\Big]
- 4 \pi F_{\mu \nu}F^{\mu \nu}\Bigg].
\eea
 Here, $\alpha_{GB}\geq 0$ is the Gauss-Bonnet coupling constant, $F^{\mu \nu} = \partial^\mu A^\nu - \partial^\nu A^\mu$ is the electromagnetic field tensor with vector potential $A^\mu$,  $\Lambda$ denotes the cosmological constant and is related to the black hole pressure by $\Lambda = -8\pi P$. In this paper we consider $D\geq 5$ since the higher curvature terms do not contribute to the dynamics for $D=4$. The theory admits spherically symmetric black hole solutions with the metric,
\be \label{spherical}
ds^2 = -f(r) dt^2 + \frac{1}{f(r)} dr^2 + r^2 h_{ij} dx^i dx^j,
\ee
where, $h_{ij} dx^i dx^j$ is the line element on the $(D-2)$ dimensional Einstein space with constant positive curvature and volume $\omega_{D-2}$. The metric function $f(r)$ is given by \cite{Cvetic:2001bk},
\begin{widetext}
\bea
f(r) = 1+ \frac{r^2}{2 \alpha} \left(1-\sqrt{1+ \frac{16 \pi \alpha M}{(D-2) \omega_{D-2} r^{D-1}}- \frac{2 \alpha Q^2}{(D-2)(D-3) r^{2D-4}} - \frac{64 \pi \alpha P}{(D-1)(D-2)}}\right).
\eea
\end{widetext}
$M, Q$ denote the mass and charge of black hole respectively, and the parameter $\alpha = (D-3)( D-4) \alpha_{GB}$. To have a well-defined {solution for $M=Q=0$,} the following condition should be satisfied,
\be
0\leq \frac{64 \pi \alpha P}{(D-1)(D-2)}\le 1.
\ee
 The mass, temperature $(T) $, and entropy of the hole can be easily obtained using the metric function as \cite{Wei:2014hba, Aranguiz:2015voa},
 \begin{widetext}
\bea
M &=& \frac{(D-2) \omega_{D-2} r_h^{D-3}}{16 \pi}\left(1+ \frac{\alpha}{r_h^2}+\frac{16 \pi P r_h^2}{(D-1)(D-2)}+\frac{\omega_{D-2} Q^2}{8\pi (D-3) r_h^{D-3}}\right), \\
 T &=& \frac{1}{4\pi r_h\left(r_h^2+2\alpha\right)}\left(\frac{16 \pi P r_h^4}{D-2} + (D-3)r_h^2+(D-5)\alpha -\frac{2 Q^2}{(D-2) r_h^{2D-8}}\right), \label{T1} \\
 S &=& \frac{\omega_{D-2} r_h^{D-2}}{4}\left(1+\frac{2(D-2)\alpha}{(D-4) r_h^2}\right).
  \eea
 \end{widetext}
Further, the equation of state is obtained as,
\bea \nonumber
P=\frac{(D-2)}{4 r_h}\left(1+\frac{2\alpha}{r_h^2}\right) T - \frac{(D-2)(D-3)}{16 \pi r_h^2}\\  \label{eos}
-     \frac{(D-2)(D-5)\alpha}{16 \pi r_h^4}+\frac{Q^2}{8\pi r_h^{2D-4}}. \label{PV}
\eea   
As mentioned before, the phase transition and critical behaviour are important aspects of black hole thermodynamics. For a canonical ensemble with fixed charge $Q$, the critical point is determined by the following condition:
\be
\frac{\partial P}{\partial r_h}\Bigg|_{r_h={r_{h_c}^*} ; T =T_c} = \frac{\partial^2 P}{\partial r_h^2}\Bigg|_{r_h={r_{h_c}^*} ; T =T_c}  =0,
\ee
where $r_{h_c}^* , T_c$ denote the critical values of horizon radius and temperature respectively. An expression for critical temperature is obtained from Eq. \ref{PV} as,
\be
T_c = \frac{10 \alpha (D-5)+ 3 (D-3) {r_{h_c}^* }^2}{4 \pi r_{h_c}^*  \left({r_{h_c}^* }^2+12 \alpha\right)}.
\ee
The existence of phase transition depends on the value of critical temperature. For example, in the case of neutral black holes in AdS space, there is a minimum temperature above which a pair of black holes can be found \cite{Hawking:1982dh}. Also, a first order phase transition is observed between these two black holes. However, phase transition for charged black holes arises only below the critical temperature\cite{Kubiznak:2012wp}. The phase transition and critical behaviour of different black hole solutions are studied using various techniques \cite{Stephens:2001sd,Carlip:2003ne, Gunasekaran:2012dq,Hendi:2012um, Mandal:2016anc, Cai:2013qga,Wei:2014hba}. This paper intends to address the black hole phase transition of {EGB} gravity in terms of topology of thermodynamic parameter space at fixed temperatures.\\

\section{Topology of Black Hole Solutions}\label{TP}
{To understand blackhole phase transition in terms of topological quantities, we adopt Duan's $\phi-$mapping topological current theory\cite{Duan:1979ucg, Duan:1984ws}. Following \cite{Fan:2022bsq}, the two-component vector field $\phi^a = (\phi^1,\phi^2)$ is defined in the space $x^\mu = (\tau, x^1, x^2)$ as,}
\be \label{phi_one}
\phi^1 = -\frac{\partial \mathcal{E}}{\partial r_h}; \qquad \phi^2 = - \cot \Theta \ \csc \Theta.
\ee

{Here, we have taken the Duan's potential to be the off-shell internal energy ($\mathcal{E}$) of the thermodynamics system. The parameter $\Theta \ (0\leq \Theta \leq \pi)$ is used for convenience \cite{Cunha:2020azh}.The location of defects are obtained from the condition given by,}
 \be
 \phi^a(\tau, \vec{x})|_{\vec{x} = \vec{z}} = 0,
 \ee
{where $\vec{z}$ denote the locations of the defects.  To study black hole phase transition which is characterized by the order parameter $r_h$ we treat the coordinates $x^1=r_h$ and $x^2=\Theta$.  The off-shell internal energy obtained from the Euclidean gravitational action is \cite{Fan:2022bsq},}
\be \label{gene_int}
\mathcal{E} = F + \frac{\tau}{\beta}.
\ee
Here $F$ denotes generalized Gibbs free energy and $\beta = T^{-1}$. {Note that the generalized internal energy is on-shell when the time coordinate $\tau$ becomes black hole entropy}. Following Duan's- $\phi$ mapping theory, a topological current associated with the vector field is constructed as \cite{Duan:1998kw,Duan:1979ucg},
\be
j^\mu = \frac{1}{2\pi} \epsilon^{\mu \nu \rho} \epsilon_{ab} \partial_\nu n^a \partial_\rho n^b; \quad \mu, \nu, \rho = 0, 1, 2; \quad a, b=1,2.
\ee
Here $\partial_\mu = \frac{\partial}{\partial x^\mu}$ and, the unit vector $n^a$ is defined by,
\be
n^a = \frac{\phi^a}{||\phi||}; \qquad ||\phi|| = \sqrt{\phi^a \phi^a}.
\ee
One can easily show that the topological current is conserved, i.e., $\partial_\mu j^\mu=0$. consequently,  a topological number is defined for a region in the parameter space $\mathcal{V}$ as,
\be\label{topnumber}
W = \int_{\mathcal{V}} d^2x j^0.
\ee
Further, the topological current can be expressed in terms of delta function \cite{Schwarz:1994ee},
\be \label{delta_rep}
j^\mu = \delta^2(\vec{\phi}) J^\mu\left(\frac{\phi}{x}\right),
\ee
where the vector Jacobian is defined by,

\be \label{Jacob}
\epsilon^{ab}J^\mu\left(\frac{\phi}{x}\right) = \epsilon^{\mu \nu \rho} \partial_\nu \phi^a \partial_\rho \phi^b.
\ee

In the topological current theory, the {zero} points of the vector field $\phi$ are treated as the topological defects. Suppose there are $N$ isolated points ($z_i; \ i=1, 2,...N$) at which $\phi=0$. The topological current constructed from the unit vector field is non-zero only at these singular points according to Eq. \ref{delta_rep}. The singular points of the vector field can be used to deduce the global topological nature of the parameter space. Using the implicit function theorem, the topological number defined in Eq. \ref{topnumber} can be expressed as,
\be
W = \int_{\mathcal{V}} d^2x j^0 = \sum_{i=1}^N \beta_i \eta_i =  \sum_{i=1}^N w_i,
\ee
where $\beta_i$ denotes the Hopf index and $\eta_i = \text{sign}\left(J^0(\frac{\phi}{x})_{z_i}\right) = \pm 1$ represents the Brouwer degree. The parameter $w_i = \beta_i \eta_i$ is called the winding number for the $i-$the singular point of the vector field $\phi$ in the parameters space $\mathcal{V}$.\\

The quantities $w_i$ and $W$ reflect the local and global topological nature respectively. A positive/negative winding number corresponds to locally thermodynamically stable/unstable system. On the other hand, topological number characterizes the global topological nature of the parameter space. As mentioned before, there are only three classes of black holes based on their topological number ($W=-1, 0, +1$) \cite{Wei:2022dzw}.\\

The kinetics of thermodynamic defects can be used to understand the nature of phase transitions. Suppose the location of $i-$th defect be at $\vec{x} = z_i(\tau)$. The velocity at which it moves is calculated by,
\be \label{velocity}
u^\mu = \frac{dz_i^\mu}{d\tau} = \frac{J^\mu(\frac{\phi}{x})}{J^0(\frac{\phi}{x})}|_{x=z_i}.
\ee
Note that $u^0=1$. In the $\phi-$ mapping theory, the zeros of the vector field at $\vec{x} = z_i$ are called regular points of $\phi$ if the Jacobian
\be
J^0\left(\frac{\phi}{x}\right) \neq 0; \qquad i=1, 2, ..N.
\ee
These defects move with a non-zero velocity in the parameters space and represent black holes. However, if,
\be
J^0\left(\frac{\phi}{x}\right)  =0  \qquad \text{and} \qquad J^1\left(\frac{\phi}{x}\right)=0, 
\ee
at some point $(\tau_*, z_*^i)$ the defect will bifurcate. Such singular points of the vector field {are} called {bifurcation points}. Existence of {bifurcation points} is crucial in our discussion of phase transition. {The velocity of defects is undetermined as the functional relation between $\tau$ and $x^1$ around $(\tau_*, z_*^i)$ becomes ill-defined at these points}. However, one can still determine its velocity by Taylor expansion of $\phi(\tau, x^1)$ around $(\tau_*, z_*^i)$ up to the order according to degeneracy of the bifurcation point (or the rank of the Jacobian matrix defined in Eq. \ref{Jacob}). If the rank is two, the velocity of the defects are obtained from the second order Taylor expansion of the field yielding\cite{Duan:1998kw, Mo:2008zz},
\be \label{const1}
\small
A \left(\frac{dx^1}{d\tau}\right)^2 + 2 B \left(\frac{dx^1}{d\tau}\right) + C =0,
\ee
where,
\be \label{const11}
\small
A = \frac{\partial^2\phi^1}{\partial (x^1)^2}\Bigg|_{\tau_*, z_*^1}; \quad B=\frac{\partial^2\phi^1}{\partial x^1 \partial \tau}\Bigg|_{\tau_*, z_*^1}; \quad C = \frac{\partial^2\phi^1}{\partial \tau^2}\Bigg|_{\tau_*, z_*^1}.
\ee
The velocity of exotic defects corresponding to bifurcation points of higher degeneracy is obtained from the cubic expansion of $\phi^1$ around  $(\tau_*, z_*^i)$. In such cases we obtain the following condition.
\be \label{const2}
\small
A_1 \left(\frac{dx^1}{d\tau}\right)^3+ 3 A_2 \left(\frac{dx^1}{d\tau}\right)^2 + 3 A_3 \left(\frac{dx^1}{d\tau}\right) + A_4=0.
\ee
where the constants are defined by,
\begin{widetext}
\be
\small
A_1 = \frac{\partial^3\phi^1}{\partial (x^1)^3}\bigg|_{\tau_*, z_*^1}; \quad A_2 = \frac{\partial^3\phi^1}{\partial (x^1)^2 \partial \tau}\bigg|_{\tau_*, z_*^1};\quad  A_3 = \frac{\partial^3\phi^1}{\partial x^1 \partial \tau^2}\bigg|_{\tau_*, z_*^1}; \quad A_4 = \frac{\partial^3\phi^1}{\partial \tau^3}\bigg|_{\tau_*, z_*^1}.
\ee
\end{widetext}
The nature of evolution of the defects can be depends on the value of the constants in Eq. \ref{const1} and Eq. \ref{const2}. The kinetics of topological defects are directly linked to the first order phase transition. The rest of the paper discusses phase transition of {EGB} black holes in AdS space in terms of topological defects.\\

\section{Topology and Phase Transition of 5D Black Holes }\label{5D}
Consider a canonical ensemble consisting of various black hole states with different values of horizon radius at a fixed temperature. Following \cite{Fan:2022bsq}, we define the the off-shell internal energy for a 5D neutral {EGB} black hole in AdS space using Eq. \ref{gene_int}. 
\bea \nonumber
\mathcal{E} &=& \frac{1}{6\pi\left(2 \alpha+r_h^2\right)} \Big[3 \pi ^2 \left(6 \alpha ^2+r_h^4-3 \alpha  r_h^2\right) \\
 &-&4 \pi ^3 P r_h^4 \left(18 \alpha +r_h^2\right)+3 r_h \tau \left(1+\frac{8 \pi }{3} P r_h^2\right)\Big].
\eea
 The temperature is obtained from Eq. \ref{T1} as,
\be \label{T_GB5}
T= \frac{8 \pi  P r_h^3+3 r_h}{12 \pi  \alpha +6 \pi  r_h^2}.
      \ee
      The equation of state is described by,
      \be
      P= \frac{3}{4 r_h} \left(1+\frac{2\alpha}{r_h^2}\right) T - \frac{3}{8 \pi r_h^2}.
      \ee
Now, the component of the vector field $\phi^1$ is obtained from Eq. \ref{phi_one} as,

\bea \nonumber
\phi^1 &=& \frac{1}{6 \pi  \left(2 \alpha +r_h^2\right)^2}\Big[ 6 \alpha + 3 r_h^2\left(\frac{8\pi P}{3}(6\alpha+r_h^2-1)\right)\Big] \\
&\times& \left( {2\pi ^2} \left({r_h}^3+6 \alpha  {r_h}\right)-\tau\right).
\eea

The Jacobians are obtained from Eq. \ref{Jacob} as,
\bea
J^1 = \frac{\partial T}{\partial r_h} =\frac{ 6 \alpha + 3 r_h^2\left(\frac{8\pi P}{3}(6\alpha+r_h^2-1)\right)}{6 \pi  \left(2 \alpha +r_h^2\right)^2},
\eea
and,
\begin{widetext}
 \bea \nonumber
J^0  &=&
\frac{1}{12 \pi  \left(2 \alpha +r_h^2\right)^3} \Bigg[4 r_h \tau  (16 \pi  \alpha  P-3) \left(r_h^2-6 \alpha \right)\\ &+&\pi ^2 \left(72 \alpha ^3+24 \pi  P r_h^8+r_h^6 (208 \pi  \alpha  P-3)+6 \alpha  r_h^4 (112 \pi  \alpha  P-3)+36 \alpha ^2 r_h^2 (48 \pi  \alpha  P-5)\right)\Bigg].
\eea
\end{widetext}
The topological defects are determined from the zero points of the vector field. Clearly, there are two types of defects. The first one is a black hole described by,
\be
\tau = \frac{\pi ^2}{2} \left(r_h^3+ 6\alpha r_h\right).
\ee
{Note that the value of $\tau$ at this defect gives the entropy of EGB black hole. }The velocity of the defect is calculated using Eq. \ref{velocity} to be,
\be
u^1= \frac{2}{3 \pi ^2 \left(2 \alpha +r_h^2\right)}.
\ee
The other type of defects characterized by $J^1=0$ are called exotic defects. These defects have zero velocity and are described by,
\be \label{HorizonGB5}
r_{h_{\pm}}^2 = \frac{\left(1-16 \pi P \alpha \right) \pm\sqrt{\left(1-\frac{16 \pi P \alpha}{3}\right)\left(1-48 \pi P \alpha\right)}}{\frac{16 \pi P}{3}}.
\ee
Note that the existence of exotic defect demands,
\be
P \leq \frac{1}{48 \pi \alpha}.
\ee
Now, there exist bifurcation points where $J^0=J^1=0$. These points are described by,
\be
r_h^* = r_{h_{\pm}}; \qquad \tau^* = \frac{\pi ^2}{2} \left({r_h^*}^3+ 6\alpha r_h^*\right).
\ee
As explained before, the velocity of the defect is undetermined at the bifurcation points. Therefore, one resorts to the method of Taylor expansion. Due to higher degeneracy, the vector field is expanded to cubic order around $(\tau^*, r_h^*)$ to obtain Eq. \ref{const2}. We find the constants as,
 \be
  A_1 = \frac{3\pi}{4 \alpha}; \quad A_2 = -\frac{1}{48 \pi \alpha^2}; \quad A_3=A_4=0.
  \ee
There are three solutions for velocity. Two of them are zero corresponding to exotic defects, the non-zero velocity,
\be
  u_*^1 = \frac{1}{12 \pi^2 \alpha} = \frac{2}{3 \pi ^2 \left(2 \alpha +{r^*_h}^2\right)},
  \ee
  represents a black hole. The velocity of defect at the bifurcation point exactly matches with the velocity of black hole obtained before.
\begin{figure}
  \centering
  \includegraphics[width=180pt]{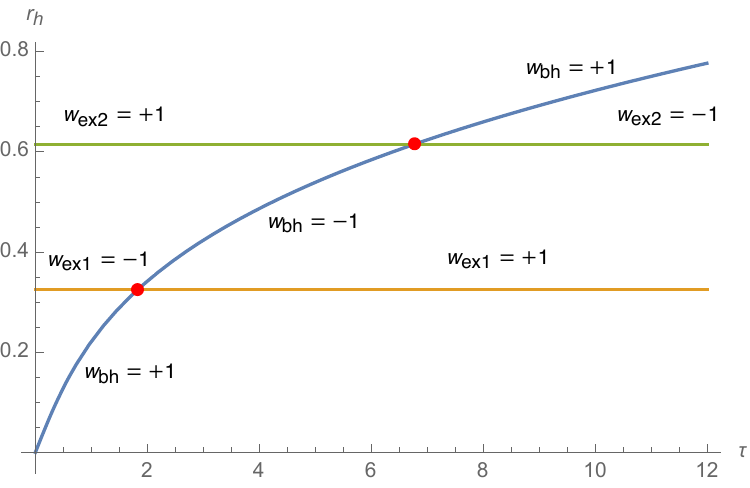}
  \caption{Defects of neutral {EGB} black hole in 5D. The blue line represents the black hole phase and the horizontal lines indicate exotic defects. The first bifurcation point at $\tau=\tau_1^*$ is labelled by the red spot between the the blue curve and the orange line. Similarly, the red spot between the blue curve and the green line at $\tau=\tau_2^*$ is the second bifurcation point.}\label{GB51}
\end{figure}

The characteristics of black hole phase transition is encoded in the evolution of topological defects. In FIG. \ref{GB51}, we have shown the locations of defects against the parameter $\tau$. A locally thermodynamically stable black hole ($w=+1$) collides with one of the static exotic defects at $\tau=\tau^*_1$. At this point, the winding number gets exchanged between the black hole and the defect. The black hole becomes unstable ($w=-1$) and continues to move, whereas the exotic defect remains static. The interchange of winding numbers is interpreted as the signature of first order phase transition. At the second bifurcation point, i.e., $\tau=\tau_2^*$, the locally thermodynamically unstable intermediate black hole collides with the second exotic defect to become thermodynamically stable ($w=+1$) large black hole phase. This interpretation corresponds to the isobar in the $P-r_h$ plot where the first order phase transition between small/large black holes take place through an isobaric process\cite{Miao:2018fke}. Note that the global topological charge is always the same ($W=+1$) due to the conservation of topological numbers.\\

\begin{figure}
  \centering
  \includegraphics[width=120pt]{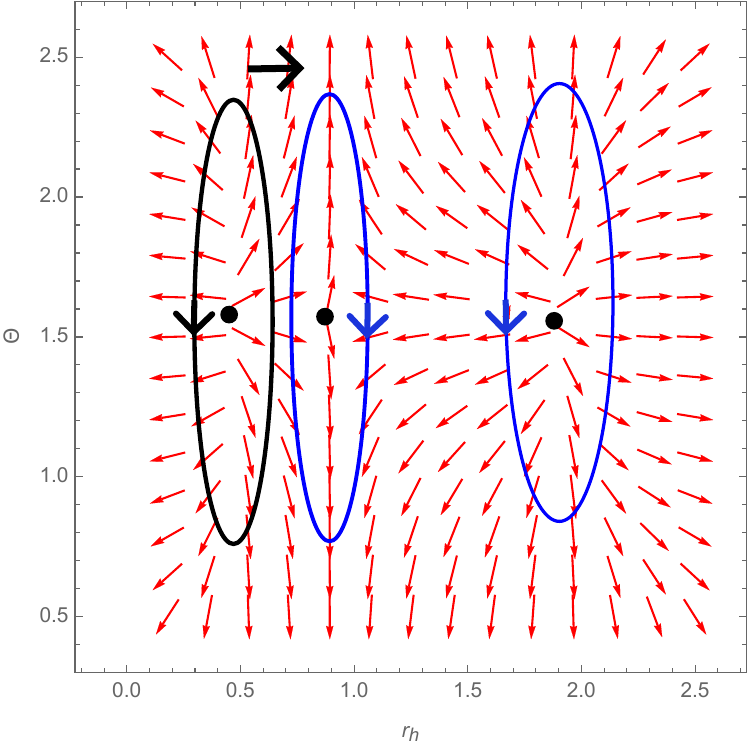}
  \includegraphics[width=120pt]{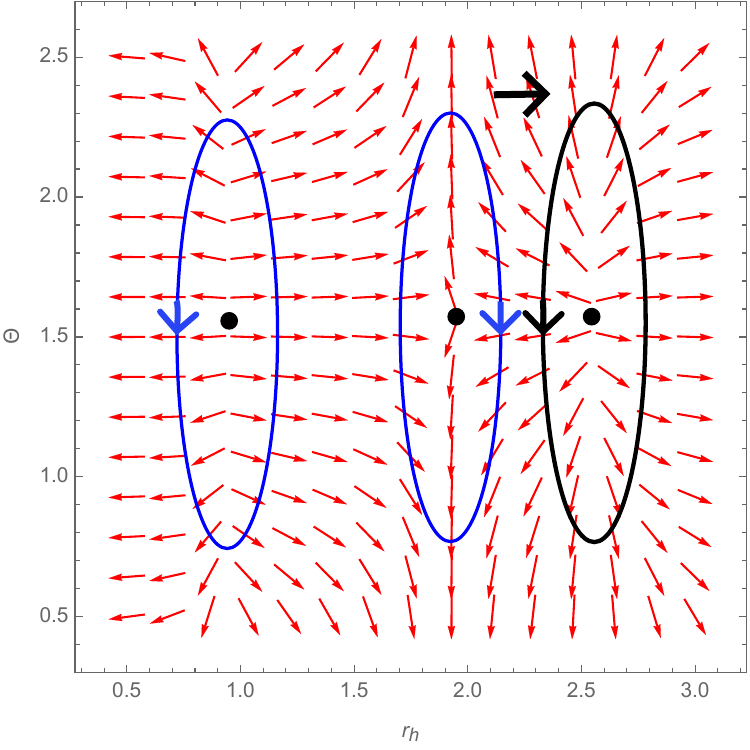}
  \caption{The interpretation of phase transition in terms of momentum transfer between black hole and exotic defects for 5D neutral black holes in {EGB} gravity. The black and blue loops represent black hole and exotic defect respectively. Clockwise and anti-clockwise direction indicate $w=-1$ and $w=+1$ respectively. \textit{On the left:} the small black hole collides with the small exotic defect elastically and exchange position and momentum. The exotic defect continue to collide with the large exotic defect elastically.  \textit{On the right:} The original large exotic defect receives momentum from the small exotic defect and becomes large black hole whereas the small exotic defect becomes large exotic defect.}
  \label{GBN}
\end{figure}

An equivalent interpretation of phase transition can be obtained from the $\phi$- field configuration. In FIG. \ref{GBN}, we have shown the vector plot between $\phi^1$ and $\phi^2$ at a fixed temperature. In this picture, an initial small black hole collides with a small exotic defect elastically with the exchange of position and momentum to become a new exotic defect. During this process the winding number of both black hole and the exotic defect remains unchanged. Now, the original small exotic defect continues to collide with the large exotic defect elastically. As a result of momentum transfer, the small exotic defect becomes a large static exotic defect whereas the original large exotic defect becomes a large black hole. This corresponds to the oscillatory region of the $P-r_h$ curve presented in \cite{Cai:2013qga,Miao:2018fke}.

Now, the critical point of the process is described by the condition where two bifurcation points meet in the parameters space, i.e., $r_{h_{+}} = r_{h_{-}} ={r_{h_c}^*}$. At the critical point, the value of horizon radius, pressure, and the temperature are given by,
\be
{r_{h_c}^*} = \sqrt{6\alpha}; \qquad P_c = \frac{1}{48 \pi \alpha}; \qquad T_c=\frac{1}{\pi \sqrt{24 \alpha}}.
\ee
The universal relation between these critical parameters is readily obtained from Eq. \ref{T_GB5} and Eq. \ref{HorizonGB5} as,
\be
\frac{P_c {r_{h_c}^*}}{T_c} = \frac{1}{4}.
\ee
Our results match with \cite{Cai:2013qga}, where authors studied phase transition using $P-V$ criticality. The phase transition of charged black holes in 5D {EGB} theory is analysed and found to be qualitatively similar to the neutral case and therefore we skip the details. As observed before, this behaviour is due to the presence of Gauss-Bonnet coupling constant $\alpha$ which behaves like charge \cite{Cai:2013qga}.\\

We find that the characteristics of phase transition for $5D$ {EGB}AdS black hole and arbitrary dimensional charged  black holes in AdS space are the same \ref{app1}. This result can be attributed to the fact that both these black hole solutions share the same topology class.\\

\section{Topology and Phase Transition in $D\geq 6$} \label{HD}
Thermodynamic properties of black holes in $D\geq6$ {EGB} gravity theory is different from the case of $D=5$. This variation can be readily observed from the equation of state given in Eq.  \ref{eos} \cite{Cai:2013qga,Wei:2014hba}. Therefore, black hole solutions in $D\geq 6$ need to be considered separately.  In this section, we address black hole phase transition in $D=6$ using topological current theory developed earlier. The vector field constructed from the generalized internal energy is obtained as,
\bea \nonumber
\phi^1&=&\frac{1}{12 \pi  \left(r_h^3+2 \alpha  r_h\right)^2}\Bigg[ \left(2 \pi ^2 \left(r_h^4+4 \alpha  r_h^2\right)-3 \tau \right) \\
&\times & \left(-2 \alpha ^2+4 \pi  P r_h^6+3 r_h^4 (8 \pi  \alpha  P-1)+3 \alpha  r_h^2\right)  \Bigg].
\eea 
The Jacobians are obtained from Eq. \ref{Jacob}.
\be
J^1= \frac{-2 \alpha ^2+4 \pi  P r_h^6+3 r_h^4 (8 \pi  \alpha  P-1)+3 \alpha  r_h^2}{4 \pi  \left(r_h^3+2 \alpha  r_h\right)^2}.
\ee
\begin{widetext}
\bea
J^0 &=& \frac{1}{6 \pi  \left(r_h^3+2 \alpha  r_h\right)^3} \Bigg[3 \tau  \left(-4 \alpha ^3+r_h^6 (8 \pi  \alpha  P-3)+12 \alpha  r_h^4 (1-4 \pi  \alpha  P)-6 \alpha ^2 r_h^2\right)\\
&+&2 \pi ^2 r_h^4 \left(36 \alpha ^3+8 \pi  P r_h^8+r_h^6 (72 \pi  \alpha  P-3)+6 \alpha  r_h^4 (40 \pi  \alpha  P-3)+2 \alpha ^2 r_h^2 (192 \pi  \alpha  P-23)\right)\Bigg].
\eea
\end{widetext}
The zero points of vector field yields two types of defects, i.e., a black hole and an exotic defect. The black hole moves with velocity,
\be
u^1 = \frac{3}{8 \pi ^2 \left(r_h^3+2 \alpha  r_h\right)},
\ee
and is described by,
\be
\tau = \frac{2}{3} \pi ^2 \left(r_h^4+4 \alpha  r_h^2\right).
\ee
The exotic defect is obtained from $J^1=0$, and is described by,
\begin{small}
\be
r_h^2=\frac{1}{4 \pi  P P_\alpha}\Bigg[1+64 \pi ^2 \alpha ^2 P^2+P_\alpha(1-8\pi \alpha P)+P_\alpha^2-20 \pi P \alpha\Bigg],
\ee
\end{small}
where,
\be
\small
P_\alpha =\Bigg[ (1-16 \pi  \alpha  P) \left(2 \pi  \alpha  P \left(16 \pi  \alpha  P+\sqrt{5-16 \pi  \alpha  P}-7\right)+1\right)\Bigg]^{\frac{1}{3}}.
\ee
In the case of six dimensions, the exotic defect is present only if,
\be
P \leq \frac{1}{16 \pi  \alpha }.
\ee
 Unlike the case of 5D, the number of exotic defects in $D= 6$ is just one. The location ($\tau^*,r^*_h$) at which the defect and black hole meet is called the bifurcation point described by $J^1=J^0=0$. The velocity of defects at this point becomes undefined. A natural way to circumvent such a situation is the Taylor expansion. Consequently, using Eq. \ref{const1}, the velocity at the bifurcation point is found to be $u_*^1=0$ or,
\be
u_*^1 = \frac{3}{8 \pi ^2 \left({r^*_h}^3+2 \alpha  {r^*_h}\right)}.
\ee

\begin{figure}
  \centering
  \includegraphics[width=180pt]{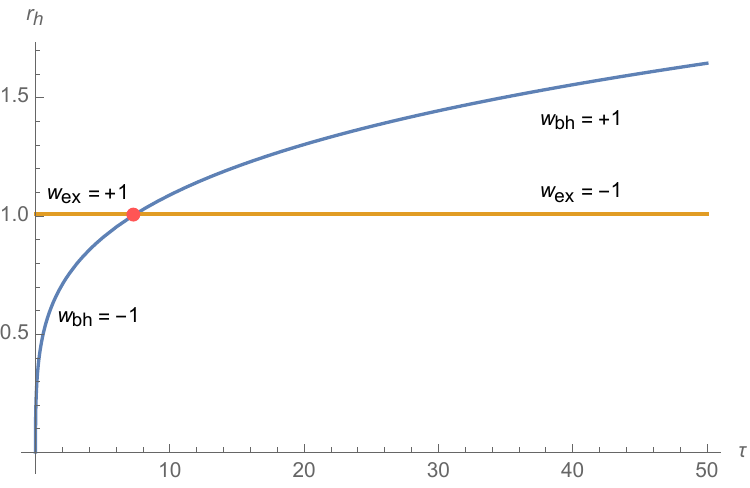}
  \caption{Defects of neutral black holes in 6D {EGB} theory in AdS space. Blue curve and horizontal line represent black hole and exotic defect respectively. At the bifurcation point $\tau=\tau^*$, black hole collides with the exotic defect with exchange of winding number.}\label{GB61}
\end{figure}

\begin{figure}
  \centering
  \includegraphics[width=120pt]{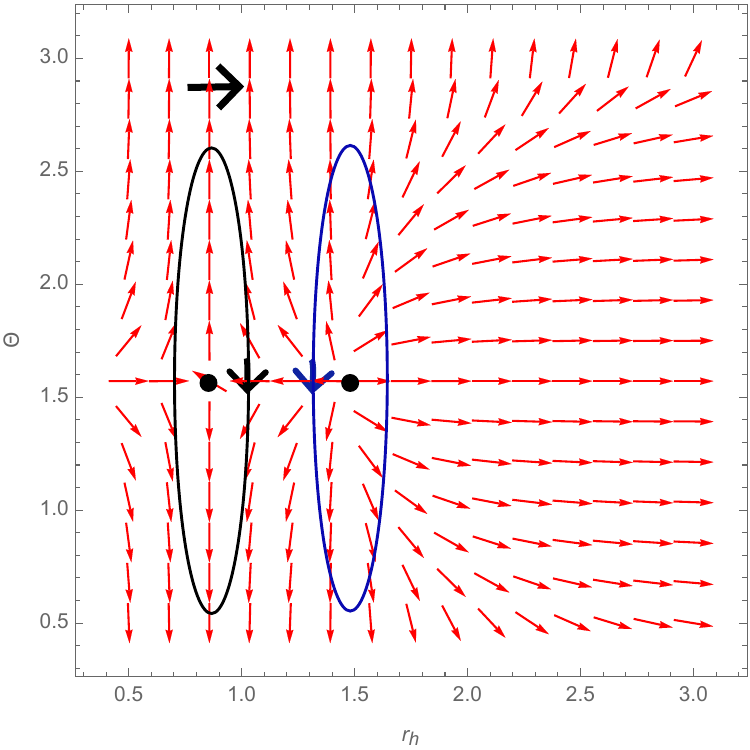}
  \includegraphics[width=120pt]{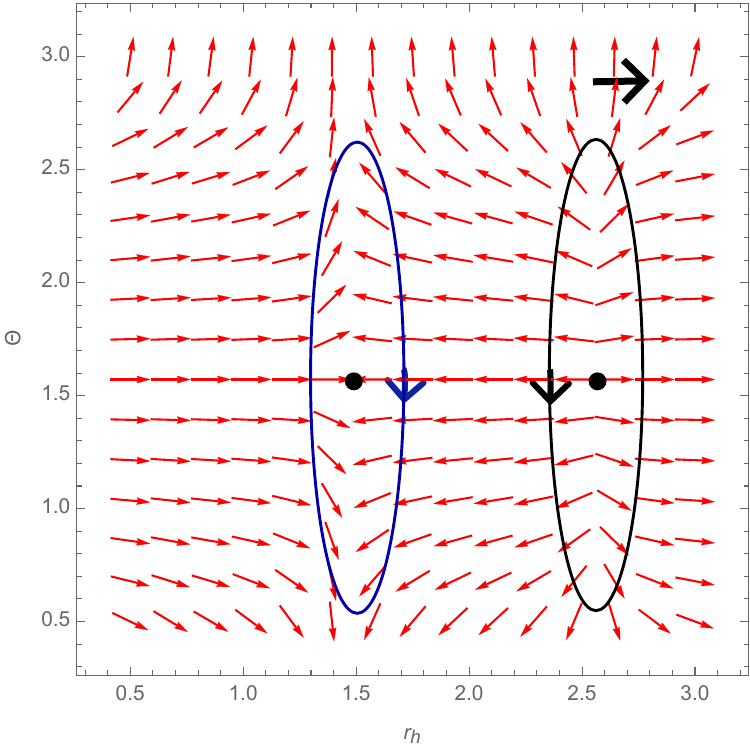}
  \caption{The defects are highlighted in the vector plot of $\phi$ fields. The black and blue loops correspond black hole and exotic defect respectively. The direction of loop indicates winding number; clockwise and anti-clockwise direction indicate $w=-1$ and $w=+1$ respectively. \textit{On the left:} the small unstable black hole collides with a static exotic defect elastically with the exchange of position and momentum. \textit{On the right:} The original exotic defect becomes a large black hole and the small black hole reduces to a static exotic defect. }
  \label{GBN6}
\end{figure}

 The non-zero velocity obtained is exactly equal to the velocity of the black hole. This shows the black hole collides with the exotic defect and departs without any change in its velocity. The defects corresponding to 6D neutral {EGB} theory are shown in FIG. \ref{GB61}. For $\tau< \tau^*$, the small black hole has a winding number $w=-1$ and the exotic defect has $w=+1$. The locally thermodynamically unstable ($w=-1$) small black hole collides with the static exotic defect at $\tau=\tau^*$. After collision, the black hole continues to move with the same velocity but with $w=+1$. Thus, the black hole becomes stable. This. characteristics, ie., the exchange of winding numbers is the signature of phase transition. Now, the exotic defect continues to stay at rest with winding number $w=-1$. For $\tau >\tau^*$, the large black hole has a winding number $w=+1$ and the exotic defect has $w=-1$. Therefore, the total topological number at any instance is the same, i.e., $W=+1-1=0$.\\
The black hole phase transition can be interpreted locally using the $\phi$- field configuration as shown in FIG. \ref{GBN6}. At constant temperature, the small black hole collides with the static exotic defect elastically with the exchange of position and momentum to become a static exotic defect. The original exotic defect evolved to become a large black hole. Note that the winding number of the black hole and the exotic defect remains unchanged during the collision. The phase transition is characterized by the momentum transfer between the black hole and the exotic defect.\\

We observe that when $Q=0$ there can only be exotic defect ($J^1=0$) for $D\geq6$ . This characteristics is similar to the case of 4D neutral black holes in AdS space\cite{Fan:2022bsq}. Therefore, the possible transition is from a small unstable black hole phase to a large stable phase.\\

\section{Conclusions and Discussions} \label{FL}
In this paper, we have provided a topological interpretation for black hole phase transitions in Einstein-Gauss-Bonnet gravity using Duan's topological current and bifurcation theories. According to the $\phi$- mapping topological current theory, black holes can be considered as topological defects in the thermodynamic parameter space. To this extent, a vector field is constructed from the generalized free energy of the system. The zeros of the vector field yield both black hole and static exotic defects. The bifurcation point is the location at which an exotic defect collides with the black hole with an interchange of winding numbers. This behaviour characterizes the black hole phase transition. Also, one can have a similar interpretation of phase transition using the vector field configuration. In this description, a black hole phase collides with the exotic defect along with the exchange of position and momentum without changing its winding number. This process can be related to the elastic collision of identical particles in classical mechanics. However, the collision between the defects should be considered virtual since the transition happens at a fixed temperature. Note that these two interpretations of phase transitions are equivalent according to Maxwell's equal area law\cite{Miao:2018fke}.\\

 In the case of 5D black holes in {Einstein-Gauss-Bonnet gravity} theory, we observe the interchange of winding number twice, indicating a stable small/large black hole phase transition through an intermediate unstable phase. The topological nature of phase transition {of 5D  Einstein-Gauss-Bonnet AdS black holes  and RNAdS black holes in any dimension are same}. In the case of black holes in $D\geq6$ {Einstein-}Gauss-Bonnet theory, the number of exotic defects is one. Consequently, we observe phase transition between an unstable small black hole and a stable large black hole. Here, the topological behaviour is similar to the case of {neutral black holes in AdS space}. Our results indicate that the similarity in the characteristics of phase transitions between certain black hole solutions can be associated to their topological nature. Therefore, we believe a class of black hole solutions with the same value of topological number exhibit similar thermodynamic properties including phase transition. However, one has to consider other black hole solutions to further verify this claim.\\
 
 In this work we have only considered black holes with spherical horizons. One can study the phase transition of  black holes with Ricci flat and hyperbolic horizons. The second order phase transitions can also be studied using the topology current theory for different black hole solutions. We leave these for future works.\\
 
\section{Acknowledgement} 

Author thanks T. K. Safir and C.L. Ahmed Rizwan for useful discussions.

\appendix

\section{Topology and Phase Transition of RNAdS Black Holes}\label{app1}
\begin{figure}
  \centering
  \includegraphics[width=180pt]{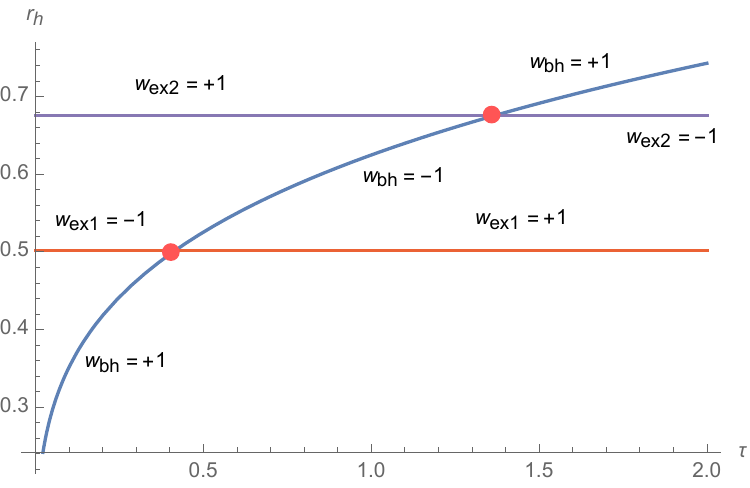}
  \caption{{Defects of RNAdS black hole in 6D. The blue line represents the black hole phase and the horizontal lines indicate exotic defects. The first bifurcation point is labelled by the red spot between the the blue curve and the orange line. Similarly, the red spot between the blue curve and the violet line is the second bifurcation point.}}\label{RNAdS_1}
\end{figure}

To justify the comparison presented in Sec. \ref{5D} between 5D Einstein-Gauss-Bonnet black holes and arbitrary dimensional RNAdS black holes, we outline the topological interpretation of RNAdS black hole phase transition using the topological current theory developed in Sec. \ref{TP}. Note that this section is a straightforward generalization of the analysis presented in \cite{Fan:2022bsq} into arbitrary dimensions. The bulk action describing the D-dimensional  Reissner-Nordstr\"{o}m black hole solution in AdS space reads,
\bea \nonumber
S=\frac{1}{16 \pi} \int d^Dx \sqrt{-g}\Big[ R - 2 \Lambda 
- 4 \pi F_{\mu \nu}F^{\mu \nu}\Big].
\eea
The spherically symmetric black hole solution can be obtained in the form of Eq. \ref{spherical} with the metric function given by,
\be
f(r)=1-\frac{16 \pi M}{(D-2) \omega_{(D-2)} r^{(D-3)}} + \frac{16 \pi^2 Q^2}{2 (D-2)(D-3)\omega^2_{(D-2)}r^{2(D-3)}} + \frac{16 \pi Pr^2}{(D-1)(D-2)}
\ee
We choose $D=6$ for the purpose of graphical analysis. The vector field constructed from the off-shell internal energy is given by,
\be
\phi^1 = \frac{1}{8\pi r_h^8}\left( 8 \pi P r_h^8 - 6 r_h^2 + 7 Q^2\right)\left(\frac{2\pi^2}{3} r_h^4-\tau\right)
\ee
The Jacobians are obtained from Eq. \ref{Jacob} as,
\be
J^1 =\frac{8 \pi P r_h^8 + 7Q^2-6r_h^6}{8 \pi r_h^6}
\ee
and,
\be
J^0 = \frac{1}{6 \pi r_h^2}\Big[ 2 \pi^2 r_h^4\left(8 \pi P r_h^8 -3 r_h^6-7 Q^2\right)+\left(42 Q^2-9 r_h^6\right)\tau\Big]
\ee
The zero points of the vector field  are called the topological defects. Two types of defects are present here, the first one represents a black hole and the second type of defects that are characterized by $J^1=0$ are called exotic defects. The bifurcation point ($J^1=J^0=0$) is identified as the location where both these defects meet. In FIG. \ref{RNAdS_1}, we have shown the locations of defects against the parameter $\tau$. The phase transition can be explained as the interchange of winding number between the black hole phase and the static exotic defect as explained in Sec. \ref{5D}.\\

\begin{figure}
  \centering
  \includegraphics[width=120pt]{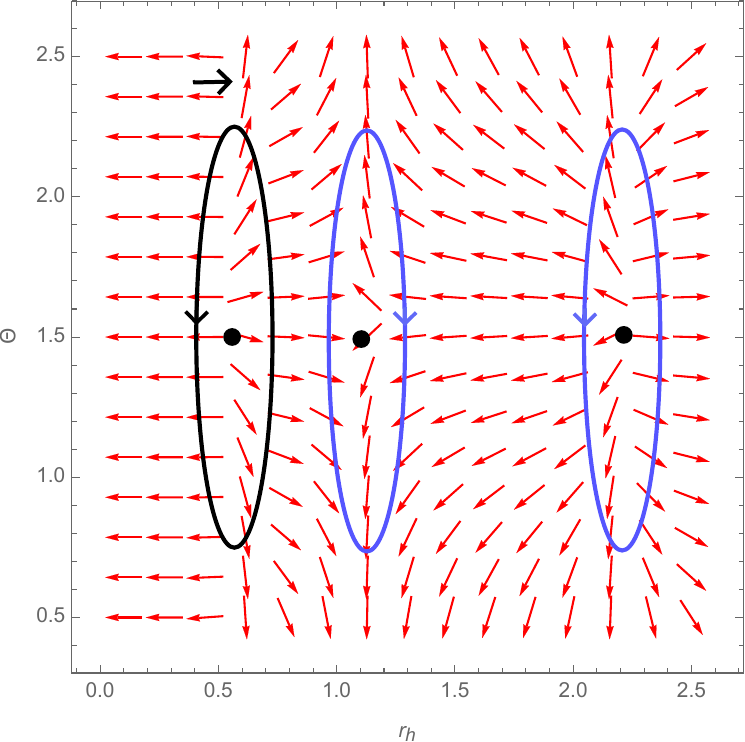}
  \includegraphics[width=120pt]{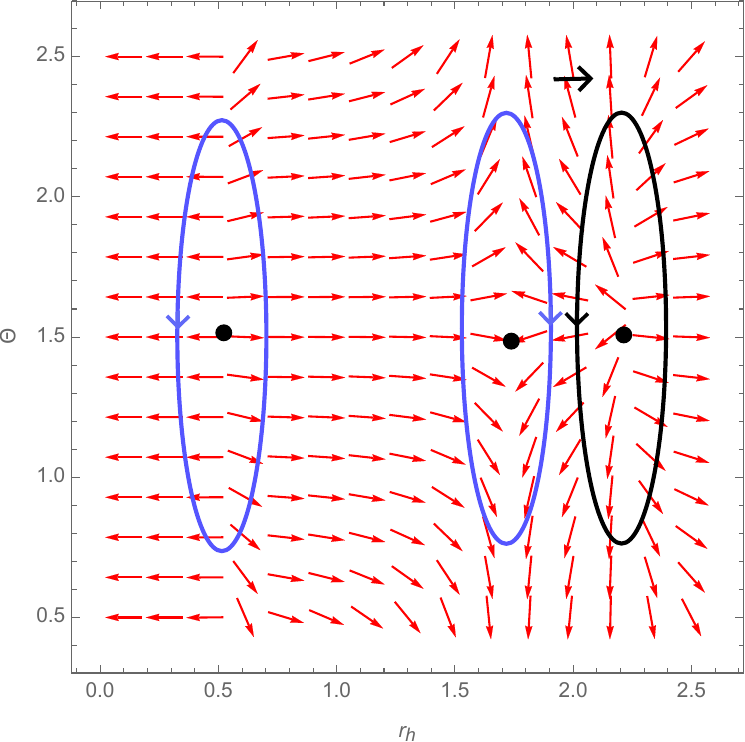}
  \caption{{The interpretation of phase transition in terms of momentum transfer between black hole and exotic defects for 6D RNAdS hole. The black and blue loops represent black hole and exotic defect respectively. Clockwise and anti-clockwise direction indicate $w=-1$ and $w=+1$ respectively. \textit{On the left:} the small black hole collides with the small exotic defect elastically and exchange position and momentum. The exotic defect continue to collide with the large exotic defect elastically.  \textit{On the right:} The original large exotic defect receives momentum from the small exotic defect and becomes large black hole whereas the small exotic defect becomes large exotic defect.}}
  \label{RNAdS_2}
\end{figure}

An equivalent interpretation of phase transition can be obtained using $\phi-$field configuration where a black hole phase collides with the exotic defect along with the exchange of position and momentum without changing its winding number.  In FIG. \ref{RNAdS_2}, we have shown the vector plot between $\phi^1$ and $\phi^2$ at a fixed temperature. Again, the interpretation of phase transition is same as given in Sec. \ref{5D}.\\

 We observe that the interpretation of phase transition of charged black holes in $D=4$ (\cite{Fan:2022bsq}) and $D=6$ AdS spaces are same. In fact, the characteristics of phase transition for charged black holes in AdS space are same for all dimensions. This can be understood on the basis of topological classification where all RNAdS black hole solutions have topological number $W=1$. Therefore, their thermodynamic features including phase transition are expected to be the same.

\end{document}